\documentclass[a4paper,twoside,11pt]{article}
\usepackage[bindingoffset=0.5cm,textheight=22.5cm,hdivide={2.9cm,*,2.85cm}, vdivide={*,22cm,*}]{geometry}
\usepackage{amsmath}
\usepackage{amsfonts}
\usepackage[dvips,bookmarksnumbered=true,breaklinks=true]{hyperref}

\newcommand{\nc}{non-com\-mu\-ta\-tive}
\newcommand{\etal}{{\it et al.}}
\newcommand{\eqnref}[1]{Eqn.~(\ref{#1})}		
\newcommand{\secref}[1]{Section~\ref{#1}}		
\newcommand{\tabref}[1]{Tab.~\ref{#1}}			
\newcommand{\co}[2]{\left[#1,#2\right]}					
\newcommand{\aco}[2]{\left\{#1,#2\right\}}				
\newcommand{\starco}[2]{\left[ #1\stackrel{\star}{,}#2\right] }		
\newcommand{\var}[2]{\frac{\d #1}{\d #2}}				
\newcommand{\pa}{\partial}						
\newcommand{\diff}[2]{\frac{\pa #1}{\pa #2}}				
\newcommand{\ri}{{\rm i}}						
\renewcommand{\k}{\tilde{k}}						
\newcommand{\Dt}{\widetilde{D}}						
\newcommand{\bc}{\bar{c}}						
\newcommand{\Act}{S}
\renewcommand{\a}{\alpha}
\renewcommand{\b}{\beta}
\newcommand{\g}{\gamma}
\renewcommand{\d}{\delta}

\renewcommand{\th}{\theta}

\renewcommand{\l}{\lambda}
\newcommand{\m}{\mu}
\newcommand{\n}{\nu}

\renewcommand{\r}{\rho}
\newcommand{\s}{\sigma}

\renewcommand{\Xi}{\Xi}

\newcommand{\inv}[1]{\frac{1}{#1}}				

\newcommand{\intx}{\int\! {\rm d}^4x}				
\newcommand{\ig}{{\rm i}g}
\newcommand{\nok}{\left(k^2+\frac{a'^2}{\tilde{k}^2}\right)} 	
\newcommand{\bpsi}{\bar{\psi}}
\newcommand{\bB}{\bar{B}}
\newcommand{\oldB}{\mathcal{B}}
\title{\vspace{2cm}Improved Localization of a Renormalizable Non-Commutative Translation Invariant U(1) Gauge Model}
\author{Daniel N. Blaschke\footnotemark[1]~,
Arnold Rofner\footnotemark[1]~, Manfred Schweda\footnotemark[1]~,\\*[12pt] and Ren\'e I.P. Sedmik\footnotemark[1]}
\date{\today}
\begin{document}
\maketitle
\thispagestyle{empty}
\begin{center}
\renewcommand{\thefootnote}{\fnsymbol{footnote}}
\vspace{-0.3cm}\footnotemark[1]Institute for Theoretical Physics,
Vienna University of Technology\\
Wiedner Hauptstrasse 8-10, A-1040 Vienna (Austria)\\[0.3cm]
\ttfamily{E-mail: blaschke@hep.itp.tuwien.ac.at,
arofner@hep.itp.tuwien.ac.at, mschweda@tph.tuwien.ac.at, sedmik@hep.itp.tuwien.ac.at}
\vspace{0.5cm}
\end{center}%
\begin{abstract}
Motivated by the recent work of Vilar \etal~\cite{Vilar:2009} we enhance our \nc~translation invariant gauge model \cite{Blaschke:2009a} by introducing auxiliary fields and ghosts forming a BRST doublet structure. In this way localization of the problematic $\inv{D^2}$ term can be achieved without the necessity for any additional degrees of freedom. The resulting theory is suspected to be renormalizable. A rigorous proof, however, has not been accomplished up to now.
\end{abstract}

%
%
\section{Introduction}

Tackling the infamous UV/IR mixing problem plaguing Moyal-deformed QFTs has been one of the main research interests in the field for almost a decade (see \cite{Tanasa:2008d,Rivasseau:2007a,Douglas:2001} for reviews of the topic.) It is accepted on a broad basis that non-commutativity creates the need for additional terms in the action to reobtain renormalizability. Several interesting approaches have been worked out \cite{Grosse:2003,Grosse:2008a}, and proofs of renormalizability have been achieved mainly by utilizing multi-scale analysis~\cite{Rivasseau:2005a,Rivasseau:2005b}, or formally in the matrix base~\cite{Grosse:2004b}.\\
Quite independent of these developments Gurau \etal~\cite{Gurau:2009} introduced a term of type $\phi\star\inv{\square}\phi$ in the Lagrangian which modifies the theory in the infrared region and, in this way, renders it renormalizable. This was in fact proven to all orders by the authors using multi-scale analysis. Motivated by the inherent translation invariance and simplicity of this model (referred to as $\inv{p^2}$ model,) a thorough study of the divergence structure and explicit renormalization at one-loop level~\cite{Blaschke:2008b} as well as a computation of the beta functions~\cite{Tanasa:2008a} have been carried out.

In the following we will premise Euclidean $\mathbb{R}^4$ with the Moyal-deformed product $\starco{x^\m}{x^\n} \equiv  x^{\mu} \star
x^{\nu} -x^{\nu} \star x^{\mu} = \ri \th^{\mu \nu}$ of regular commuting coordinates $x^\mu$. The real parameters $\th^{\m\n}=-\th^{\n\m}$ form the block-diagonal tensor
\begin{align}\label{eq:def-theta}
( \th^{\mu\nu} )
=\th\left(\begin{array}{cccc}
0&1&0&0\\
-1&0&0&0\\
0&0&0&1\\
0&0&-1&0
\end{array}\right)\, ,  \quad \text{with } \; \th \in \mathbb{R} \, .
\end{align}
With these definitions we use the abbreviations $\tilde{v}^\m \equiv \th^{\m\n}v_\n$ for vectors $v$ and $\tilde{M} \equiv \th^{\m\n}M_{\m\n}$ for matrices $M$.\\[2ex]
\indent A generalization of the scalar $\phi^4$, $\inv{p^2}$ model to U(1) gauge theory was first proposed in~\cite{Blaschke:2008a} yielding the action
\begin{align}
 \hspace{-6pt}\Act&=\Act_{\text{inv}}[A]+\Act_{\text{gf}}[A,b,c,\bc]\nonumber\\
 \hspace{-6pt}&=\!\intx\Big[\inv{4}F_{\m\n}\star F^{\m\n}\!+F_{\m\n}\star\inv{D^2\Dt^2}\star F^{\m\n}\Big]+\!\intx \Big[b\star\partial\cdot A-\frac{\alpha}{2}b\star b - \bc\star\partial^\m D_\m c\Big],
 \label{eq:act_old_complete}
\end{align}
with the usual gauge boson $A_\m$, ghost and antighost fields $c$ and $\bc$ respectively, the Lagrange multiplier field $b$, and a real U(1) gauge parameter $\alpha$. The antisymmetric field strength tensor $F_{\m\n}$ and the covariant derivative $D_\m$ are defined by
\begin{align}
 F_{\m\n}=\partial_\m A_\n -\partial_\n A_\m -\ri g \starco{A_{\m}}{A_{\n}}, \quad \text{and } D_{\m}\varphi= \partial_\m\varphi-\ri g \starco{A_\m}{\varphi},
\label{eq:act_damping_term}
\end{align}
for arbitrary $\varphi$. The term 
\begin{align}
\Act_{\text{nloc}}=\intx\, F_{\m\n}\star\inv{D^2\Dt^2}\star F^{\m\n},
\end{align}
implements the damping mechanism of the $\inv{p^2}$ model by Gurau~\etal~\cite{Gurau:2009}. It is of inherently non-local nature and gives rise to an infinite number of gauge boson vertices as described in Ref.~\cite{Blaschke:2008b}. 
However, there are several approaches for localization which deserve to be discussed in more detail below in \secref{sec:localization}. First, we briefly review our early attempts in this endeavour which led to the introduction of additional degrees of freedom. Subsequently, as an alternative way to tackle the problem, a modified version of the model with auxiliary fields forming BRST doublet structures is discussed. This latter approach aims to form a basis for the application of algebraic (or equivalently any other) renormalization which will be postponed to a future publication. The UV power counting is discussed in \secref{sec:power_counting}, and a short conclusion and outlook are given in \secref{sec:conclusion}.

%
\section{The Art of Localization}
%
\label{sec:localization}

Recently two different methods of localization for terms of the type $\inv{D^2}$ have been proposed \cite{Blaschke:2009a,Vilar:2009}. In the following these approaches shall be reviewed briefly.
%
\subsection{Introduction of an Auxiliary Field}
%
\label{sec:loc_old_model}
The problem of the infinite power series raised by \eqnref{eq:act_damping_term} can be avoided by the introduction of at least one additional real antisymmetric field $\oldB_{\m\n}$ of mass dimension two, as described in~\cite{Blaschke:2009a}:
\begin{align}
\Act_{\text{nloc}}\to\Act_{\text{loc}}=\intx\left[a'\oldB_{\mu\nu}\star F_{\mu\nu}-\oldB_{\mu\nu}\star \Dt^2D^2\star \oldB_{\mu\nu}\right]\,,
\label{eq:gauge_act_damping_replacement}
\end{align}
which represents the simplest case. However, this method introduces new physical degrees of freedom which can only be avoided if the new terms in the action are written as an exact BRST variation. In order for such a mechanism to work, further unphysical fields are required.

\subsection{Localization \`{a} la Vilar et al.}
%
Recently, Vilar \etal~\cite{Vilar:2009} found an alternative way to rewrite the critical term \eqref{eq:act_damping_term} by introducing two pairs of auxiliary complex conjugated antisymmetric tensorial fields $(B_{\m\n},\bar{B}_{\m\n})$, and $(\chi_{\m\n},\bar{\chi}_{\m\n})$ of mass dimension one, 
\begin{align}
\Act_{\text{nloc}}\to\Act_{\text{loc}}&=\Act_{\text{loc,0}}+\Act_{\text{break}}\nonumber\\
 &= \intx \left(\bar{\chi}_{\m\n}\star D^2B_{\m\n}+\bar{B}_{\m\n}\star D^2\chi_{\m\n} + \gamma^2\bar{\chi}_{\m\n}\star\chi_{\m\n}\right)+\Act_{\text{break}}\, ,
\label{eq:vilar_act_damping_replacement}
\end{align} 
with $\gamma$ being a parameter of mass dimension one. The term $\Act_{\text{nloc}}$ is now split into a BRST invariant part $\Act_{\text{loc,0}}$, and a breaking term $\Act_{\text{break}}$ as can be seen by explicit calculation with the definitions in Ref.~\cite{Vilar:2009}. The additional degrees of freedom are eliminated by following the ideas of Zwanziger~\cite{Zwanziger:1989} (see \cite{Dudal:2008} for a more comprehensive review of the topic) to add a ghost for each auxiliary field in such a way that BRST doublet structures are formed. This results in a trivial BRST cohomology for $\Act_{\text{loc,0}}$ from which follows~\cite{Baulieu:2009} that 
\begin{align}
s\Act_{\text{loc,0}}=0\quad \Rightarrow\quad \Act_{\text{loc,0}} = s \hat{\Act}_{\text{loc,0}}\,,
\end{align}
i.e. the part of the action depending on the auxiliary fields and their associated ghosts can be written as an exact expression with respect to the nilpotent BRST operator $s$. Unfortunately the breaking term does not join this nice property due to a non-trivial cohomology. However, it is constructed such that its mass dimension is smaller than four, the dimension of the underlying Euclidean space. Such a breaking is referred to as ``soft'' (c.f. Ref.~\cite{Baulieu:2009}) and does not spoil renormalizability. In fact, $\Act_{\text{break}}$ is the actual origin of the avoidance of UV/IR mixing featured by this theory as it alters the IR sector while not affecting the UV part. The mechanism of soft breaking in combination with UV renormalization will be discussed in the subsequent sections below.
%
\section{New Model Omitting the Double Quartet Structure}
%
\label{sec:new_model}

The starting point is the gauge invariant part $\Act_{\text{loc}}$ (c.f. \eqnref{eq:gauge_act_damping_replacement}) of the action \eqref{eq:act_old_complete} which has
originally been introduced to localize the term \eqref{eq:act_damping_term}. As has been shown in \cite{Blaschke:2009a} the auxiliary field $\oldB_{\m\n}$ appears to have its own dynamic properties representing additional degrees of freedom. Following \cite{Vilar:2009} we introduce an additional pair of ghost and antighost fields $\psi_{\m\n}$ and $\bpsi_{\m\n}$, and furthermore turn $\oldB_{\m\n}$ into a complex conjugated pair ($B_{\m\n}$, $\bB_{\m\n}$) of fields in order to replace \eqnref{eq:gauge_act_damping_replacement} with
\begin{align}\label{act-loc}
\Act_{\text{loc}}&=\intx\left[\frac{\l}{2}\left(B_{\m\n}+\bB_{\m\n}\right)F^{\m\n}-\mu^2\bB_{\m\n}D^2\Dt^2B^{\m\n}+\mu^2\bpsi_{\m\n}D^2\Dt^2\psi^{\m\n}\right],
\end{align}
where (as in the remainder of this section) all field products are considered to be star products. 
The parameters $\l$ and $\mu$ both have mass dimension 1 and replace the former dimensionless parameter $a'$.

It is easy to show the equivalence of this localized action and the original non-local version by employing the path integral formalism:
\begin{align}
Z&=\int\mathcal{D}(\bpsi\psi\bB BA)\,\exp\left\{-\left(\intx\inv{4}F_{\m\n}F^{\m\n}+\Act_{\text{loc}}\right)\right\}\nonumber\\
&=\int\mathcal{D}(\bB BA)\,{\det}^4\left(\mu^2D^2\Dt^2\right)\exp\bigg\{-\intx\bigg[\inv{4}F_{\m\n}F^{\m\n}+\frac{\l}{2}\left(B_{\m\n}+\bB_{\m\n}\right)F^{\m\n}\nonumber\\
&\hspace{7.7cm}-\mu^2\bB_{\m\n}D^2\Dt^2B^{\m\n}\bigg]\bigg\}\nonumber\\
&=\int\mathcal{D}(\bB BA)\,{\det}^4\left(\mu^2D^2\Dt^2\right)\exp\left\{-\intx\left[\inv{4}F_{\m\n}F^{\m\n}+\frac{\l^2}{4\mu^2}F_{\m\n}\inv{\Dt^2D^2}F^{\m\n}-\right.\right.\nonumber\\
&\hspace{3.6cm}\left.\left.-\left(\bB_{\m\n}-\frac{\l}{2\mu^2}\inv{\Dt^2D^2}F_{\m\n}\right)\mu^2D^2\Dt^2\left(B^{\m\n}-\frac{\l}{2\mu^2}\inv{\Dt^2D^2}F^{\m\n}\right)\right]\!\right\}\nonumber\\
&=\int\mathcal{D}A\,{\det}^4\!\left(D^2\Dt^2\right){\det}^{-4}\!\left(D^2\Dt^2\right)\exp\left\{-\intx\left[\inv{4}F_{\m\n}F^{\m\n}+\frac{\l^2}{4\m^2}F_{\m\n}\inv{\Dt^2D^2}F^{\m\n}\right]\!\right\}\!.
\end{align}

When considering Landau gauge fixing, i.e.
\begin{align}
\Act_{\phi\pi}=\intx\left(b\partial^\m A_\m-\bc\partial^\m D_\m c\right)\,,
\end{align}
one has the following BRST transformation laws for the fields:
\begin{align}
 &sA_\mu=D_\mu c\,, && sc=\ri g{c}{c}\, ,\nonumber\\
 &s\bc=b\,,                                                       && sb=0\, ,  \nonumber\\
 &sF_{\m\n}=\ri g\co{c}{F_{\m\n}}\,,\label{eq:BRST_standard_fields}
\intertext{and furthermore}
& s\bpsi_{\mu\nu}=\bB_{\m\n}+\ri g\aco{c}{\bpsi_{\m\n}}\,,                 && s\bB_{\m\n}=\ri g\co{c}{\bB_{\m\n}}\,,\nonumber\\
& sB_{\m\n}=\psi_{\m\n}+\ri g\co{c}{B_{\m\n}}\,,                    && s\psi_{\m\n}=\ri g\aco{c}{\psi_{\m\n}}\,.\label{eq:BRST_auxiliary_fields}
\end{align}
Observe that with the BRST doublet structure of \eqnref{eq:BRST_auxiliary_fields} one can write
\begin{align}\label{act-new}
\Act_{\text{loc}}=\intx\left[\frac{\l}{2}B_{\m\n}F^{\m\n}+s\left(\frac{\l}{2}\bpsi_{\m\n}F^{\m\n}-\mu^2\bpsi_{\m\n}D^2\Dt^2B^{\m\n}\right)\right],
\end{align}
where the first term gives rise to a breaking of BRST invariance, as
\begin{align}\label{eq:act_break_wo_source}
s\Act_{\text{break}}&=\intx\frac{\l}{2}\psi_{\m\n}F^{\m\n}, \quad \text{with }\quad \Act_{\text{break}}=\intx\frac{\l}{2}B_{\m\n}F^{\m\n}\,.
\end{align}
Since the mass dimension $d_m$ of the field dependent part of $\Act_{\text{break}}$ fulfills the condition $d_m\left(\psi_{\m\n}F^{\m\n}\right)=3<D=4$ the breaking is considered to be ``soft''. As has been shown by Zwanziger~\cite{Zwanziger:1993} terms of this type do not spoil renormalizability. In order to restore BRST invariance in the UV region (as is a prerequisite for a future application of algebraic renormalization) an additional set of sources 
\begin{align}
&s \bar{Q}_{\m\n\a\b}=\bar{J}_{\m\n\a\b}+\ri g \aco{c}{\bar{Q}_{\m\n\a\b}}, && s \bar{J}_{\m\n\a\b}=\ri g \co{c}{\bar{J}_{\m\n\a\b}}\,,
\nonumber\\
&s Q_{\m\n\a\b}=J_{\m\n\a\b}+\ri g \aco{c}{Q_{\m\n\a\b}}, && s J_{\m\n\a\b}=\ri g \co{c}{J_{\m\n\a\b}}\label{eq:brst_sources_q_j}\,,
\end{align}
is introduced, and coupled to the breaking term which then takes the form
\begin{align}
\Act_{\text{break}}&=\intx\,s\left(\bar{Q}_{\m\n\a\b}B^{\m\n}F^{\a\b}\right)\nonumber\\
&=\intx\,\left(\bar{J}_{\m\n\a\b}B^{\m\n}F^{\a\b}-\bar{Q}_{\m\n\a\b}\psi^{\m\n}F^{\a\b}\right)\,.
\end{align}
\eqnref{eq:act_break_wo_source} is reobtained if the sources  $\bar{Q}$ and $\bar{J}$ take their ``physical values''
\begin{align}\label{JQ-phys}
\bar{Q}_{\m\n\a\b}\big|_{\text{phys}}=0\,,\qquad \bar{J}_{\m\n\a\b}\big|_{\text{phys}}=\frac{\l}{4}\left(\d_{\m\a}\d_{\n\b}-\d_{\m\b}\d_{\n\a}\right)\,,\nonumber\\
Q_{\m\n\a\b}\big|_{\text{phys}}=0\,,\qquad J_{\m\n\a\b}\big|_{\text{phys}}=\frac{\l}{4}\left(\d_{\m\a}\d_{\n\b}-\d_{\m\b}\d_{\n\a}\right)\,.
\end{align}
Note that the hermitian conjugate of the counterterm $\Act_{\text{break}}$ in \eqnref{act-loc}, (i.e. the term $\intx\bB_{\m\n}F_{\m\n}$) may also be coupled to external sources which, however, is not required for BRST invariance but restores hermiticity of the action. 
\begin{align}
\frac{\l}{2}\intx \bB_{\m\n}F_{\m\n}\ \longrightarrow \ \intx\,s\left(J_{\m\n\a\b}\bpsi_{\m\n}F_{\a\b}\right)=\intx\,J_{\m\n\a\b}\bB_{\m\n}F_{\a\b}.
\end{align}

Including external sources $\Omega^\phi,\,\phi\in\{A,c,B,\bB,\psi,\bpsi,J,\bar{J},Q,\bar{Q}\}$ for the non-linear BRST transformations the complete action with Landau gauge $\pa^\m A_\m=0$ and general $Q/\bar{Q}$ and $J/\bar{J}$ reads:
\begin{align}\label{complete-action1}
\Act&=\Act_{\text{inv}}+\Act_{\phi\pi}+\Act_{\text{new}}+\Act_{\text{break}}+\Act_{\text{ext}}\,,
\end{align}
with
{\allowdisplaybreaks
\begin{align}
\Act_{\text{inv}}&=\intx\inv{4}F_{\m\n}F^{\m\n}\,,\nonumber\\
\Act_{\phi\pi}&=\intx\,s\left(\bc\,\pa^\m A_\m\right)=\intx\left(b\,\pa^\m A_\m-\bc\,\pa^\m D_\m c\right)\,,\nonumber\\
\Act_{\text{new}}&=\intx\,s\left(J_{\m\n\a\b}\bpsi^{\m\n}F^{\a\b}-\mu^2\bpsi_{\m\n}D^2\Dt^2B^{\m\n}\right)\nonumber\\*
&=\intx\left(J_{\m\n\a\b}\bB^{\m\n}F^{\a\b}-\mu^2\bB_{\m\n}D^2\Dt^2B^{\m\n}+\mu^2\bpsi_{\m\n}D^2\Dt^2\psi^{\m\n}\right)\,,\nonumber\\
\Act_{\text{break}}&=\intx\,s\left(\bar{Q}_{\m\n\a\b}B^{\m\n}F^{\a\b}\right)=\intx\left(\bar{J}_{\m\n\a\b}B^{\m\n}F^{\a\b}-\bar{Q}_{\m\n\a\b}\psi^{\m\n}F^{\a\b}\right)\,,\nonumber\\
\Act_{\text{ext}}&=\intx\left(\Omega^A_\m D^\m c+\ig\, \Omega^c c c+\Omega^{B}_{\m\n}\left(\psi^{\m\n}+\ig\co{c}{B^{\m\n}}\right) +\ig\,\Omega^{\bB}_{\m\n}\co{c}{\bB^{\m\n}}\right.\nonumber\\*
&\quad\left. +\ig\, \Omega^\psi_{\m\n}\aco{c}{\psi^{\m\n}}+ \Omega^{\bpsi}_{\m\n}\left(\bB^{\m\n}+\ig \aco{c}{\bpsi^{\m\n}}\right)+\Omega^{Q}_{\m\n\a\b}\left( J^{\m\n\a\b}+\ig \aco{c}{Q^{\m\n\a\b}}\right) \right.\nonumber\\*
&\quad\left. +\ig\, \Omega^J_{\m\n\a\b}\co{c}{J^{\m\n\a\b}}+\Omega^{\bar{Q}}_{\m\n\a\b}\left( \bar{J}^{\m\n\a\b}+\ig \aco{c}{\bar{Q}^{\m\n\a\b}}\right) +\ig\, \Omega^{\bar{J}}_{\m\n\a\b}\co{c}{\bar{J}^{\m\n\a\b}}\right).
\label{eq:act_complete}
\end{align}}
\tabref{tab:field_prop} summarizes properties of the fields and sources contained in the model \eqref{eq:act_complete}.
\begin{table}[!hb]
\caption{Properties of fields and sources.}
\label{tab:field_prop}
\begin{tabular}{l c c c c c c c c c c c}
\hline
\hline
\rule[12pt]{0pt}{0.1pt}
Field       & $A_\m$ & $c$ & $\bc$ & $B_{\m\n}$ & $\bB_{\m\n}$ & $\psi_{\m\n}$ & $\bpsi_{\m\n}$ & $J_{\a\b\m\n}$ & $\bar{J}_{\a\b\m\n}$ & $Q_{\a\b\m\n}$ & $\bar{Q}_{\a\b\m\n}$\\[2pt]
\hline
$g_\sharp$  &    0   &  1  &   -1  &     0      &    0         &        1      &      -1        &   0            &       0              &  -1            &  -1\\
Mass dim.   &    1   &  0  &   2   &     1      &    1         &        1      &       1        &   1            &       1              &  1             &  1\\
Statistics  &    b   &  f  &   f   &     b      &    b         &        f      &       f        &   b            &       b              &  f             &  f\\
\hline
\rule[14pt]{0pt}{0.1pt}
Source      & $\Omega^A_\m$ & $\Omega^c$ & $b$ & $\Omega^B_{\m\n}$ & $\Omega^{\bB}_{\m\n}$ & $\Omega^{\psi}_{\m\n}$ & $\Omega^{\bpsi}_{\m\n}$ & $\Omega^J_{\a\b\m\n}$ & $\Omega^{\bar{J}}_{\a\b\m\n}$ &$\Omega^Q_{\a\b\m\n}$ & $\Omega^{\bar{Q}}_{\a\b\m\n}$\\[2pt]
\hline
$g_\sharp$  &   -1   &  -2 &   0   &    -1      &   -1         &       -2      &       0        &  -1            &      -1              & 0             &  0\\
Mass dim.   &    3   &  4  &   2   &     3      &    3         &        3      &       3        &   3            &       3              &  3            &  3\\
Statistics  &    f   &  b  &   b   &     f      &    f         &        b      &       b        &   f            &       f              &  b            &  b\\
\hline
\hline
\end{tabular}
\end{table}

Notice that the mass $\mu$ is a physical parameter despite the fact that the variation of the action $\diff{\Act}{\mu^2}=s\left(\bpsi_{\m\n}D^2\Dt^2B^{\m\n}\right)$ yields an exact BRST form. Following the argumentation in Ref.~\cite{Baulieu:2009} this is a consequence of the introduction of a soft breaking term. For vanishing Gribov-like parameter $\lambda$ the contributions to the path integral of the $\mu$ dependent sectors of $\Act_{\text{new}}$ in \eqref{eq:act_complete} cancel each other. If $\lambda\neq0$ one has to consider the additional breaking term which couples the gauge field $A_\mu$ to the auxiliary field $B_{\mu\nu}$ and the associated ghost $\psi_{\m\n}$. This mixing is reflected by the appearance of $a'=\lambda/\mu$ in the damping factor $\nok$ featured by all field propagators \eqref{eq:prop_aa}--\eqref{eq:prop_bb}.
%
\subsection{Symmetries}
%
Aiming to apply the method of algebraic renormalization we explore the symmetry content of the proposed theory. 
The Slavnov-Taylor identity is given by
\begin{align}\label{mod2-SL-id}
 \mathcal{B}(\Act)=&\intx\Big[\var{\Act}{\Omega^A_\m}\var{\Act}{A^\m}+\var{\Act}{\Omega^c}\var{\Act}{c}+ b \var{\Act}{\bc}+\var{\Act}{\Omega^B_{\m\n}}\var{\Act}{B^{\m\n}}+\var{\Act}{\Omega^{\bB}_{\m\n}}\var{\Act}{\bB^{\m\n}}\nonumber\\
&\phantom{\intx\Big[}+\var{\Act}{\Omega^\psi_{\m\n}}\var{\Act}{\psi^{\m\n}}+\var{\Act}{\Omega^{\bpsi}_{\m\n}}\var{\Act}{\bpsi^{\m\n}}+\var{\Act}{\Omega^{Q}_{\m\n\a\b}}\var{\Act}{Q^{\m\n\a\b}}+\var{\Act}{\Omega^J_{\m\n\a\b}}\var{\Act}{J^{\m\n\a\b}}\nonumber\\
&\phantom{\intx\Big[}+\var{\Act}{\Omega^{\bar{Q}}_{\m\n\a\b}}\var{\Act}{\bar{Q}^{\m\n\a\b}}+\var{\Act}{\Omega^{\bar{J}}_{\m\n\a\b}}\var{\Act}{\bar{J}^{\m\n\a\b}}\Big]=0\,.
\end{align}
Furthermore we have the gauge fixing condition
\begin{align}
\var{\Act}{b}=\pa^\m A_\m=0\,,
\end{align}
the ghost equation
\begin{align}
\mathcal{G}(\Act)=\pa_\m\var{\Act}{\Omega^A_\m}+\var{\Act}{\bc}=0\,,
\end{align}
and the antighost equation
\begin{align}
\bar{\mathcal{G}}(\Act)=\intx \var{\Act}{c}=0\,.
\end{align}

Following the notation of Ref.~\cite{Vilar:2009} the identity associated to the BRST doublet structure is given by
\begin{align}
\hspace{-4pt}U^{(1)}_{\a\b\m\n}(\Act)&=\intx\left(\!\bB_{\a\b}\var{\Act}{\bpsi^{\m\n}}+\Omega^{\bpsi}_{\m\n}\var{\Act}{\Omega^{\bB}_{\a\b}}+\psi_{\a\b}\var{\Act}{B^{\m\n}}-\Omega^B_{\m\n}\var{\Act}{\Omega^{\psi}_{\a\b}}\right.\nonumber\\
\hspace{-4pt}&\quad\left.+J_{\m\n\r\s}\var{\Act}{Q^{\a\b}_{\phantom{\a\b}\r\s}}+\Omega^{Q}_{\a\b\r\s}\var{\Act}{\Omega^J_{\m\n\r\s}}+\bar{J}_{\m\n\r\s}\var{\Act}{\bar{Q}^{\a\b}_{\phantom{\a\b}\r\s}}+\Omega^{\bar{Q}}_{\a\b\r\s}\var{\Act}{\Omega^{\bar{J}}_{\m\n\r\s}}\right)\!=0\,.
\label{eq:u1_symmetry}
\end{align}
It is interesting to mention that the first two terms of the second line, 
\[\intx\left( J_{\m\n\r\s}\var{\Act}{Q^{\a\b}_{\phantom{\a\b}\r\s}}+\Omega^{Q}_{\a\b\r\s}\var{\Act}{\Omega^J_{\m\n\r\s}}\right)=0\,,\] 
constitute a symmetry by themselves. These terms stem from the insertion of conjugated field partners $J$ and $Q$ for $\bar{J}$ and $\bar{Q}$, respectively, which are not necessarily required as discussed above in \secref{sec:new_model}.

Furthermore, we have the linearly broken symmetries $U^{(0)}$ and $\tilde{U}^{(0)}$:
\begin{align}
U^{(0)}_{\a\b\m\n}(\Act)=-\Theta^{(0)}_{\a\b\m\n}=-\tilde{U}^{(0)}_{\a\b\m\n}(\Act)\,,
\end{align}
with
\begin{align}
\hspace{-4pt}U^{(0)}_{\a\b\m\n}(\Act)&=\!\intx\left[B_{\a\b}\var{\Act}{B_{\m\n}}-\bB_{\m\n}\var{\Act}{\bB_{\a\b}}-\Omega^B_{\m\n}\var{\Act}{\Omega^B_{\a\b}}+\Omega^{\bB}_{\a\b}\var{\Act}{\Omega^{\bB}_{\m\n}}\right.\nonumber\\
\hspace{-4pt}&\phantom{=\!\intx}\!\left.+J_{\a\b\r\s}\var{\Act}{J_{\m\n\r\s}}-\bar{J}_{\m\n\r\s}\var{\Act}{\bar{J}_{\a\b\r\s}}-\Omega^J_{\m\n\r\s}\var{\Act}{\Omega^J_{\a\b\r\s}}+\Omega^{\bar{J}}_{\a\b\r\s}\var{\Act}{\Omega^{\bar{J}}_{\m\n\r\s}}\right],\\
\hspace{-4pt}\tilde{U}^{(0)}_{\a\b\m\n}(\Act)&=\!\intx\left[\psi_{\a\b}\var{\Act}{\psi_{\m\n}}-\bpsi_{\m\n}\var{\Act}{\bpsi_{\a\b}}-\Omega^{\psi}_{\m\n}\var{\Act}{\Omega^{\psi}_{\a\b}}+\Omega^{\bpsi}_{\a\b}\var{\Act}{\Omega^{\bpsi}_{\m\n}}\right.\nonumber\\
\hspace{-4pt}&\phantom{=\!\intx}\!\left.+Q_{\a\b\r\s}\var{\Act}{Q_{\m\n\r\s}}-\bar{Q}_{\m\n\r\s}\var{\Act}{\bar{Q}_{\a\b\r\s}}-\Omega^Q_{\m\n\r\s}\var{\Act}{\Omega^Q_{\a\b\r\s}}+\Omega^{\bar{Q}}_{\a\b\r\s}\var{\Act}{\Omega^{\bar{Q}}_{\m\n\r\s}}\right],\\
\hspace{-4pt}\Theta^{(0)}_{\a\b\m\n}&=\!\intx\left[\bB_{\m\n}\Omega^{\bpsi}_{\a\b}-\psi_{\a\b}\Omega^B_{\m\n}+\bar{J}_{\m\n\r\s}\Omega^{\bar{Q}}_{\a\b\r\s}-J_{\a\b\r\s}\Omega^{Q}_{\m\n\r\s}\right].
\end{align}
Finally, the model features a symmetry denoted by $U^{(2)}$,
\begin{align}
 U^{(2)}_{\m\n\a\b}=\intx\left(\psi_{\m\n}\var{\Act}{\bpsi_{\a\b}}+\psi_{\a\b}\var{\Act}{\bpsi_{\m\n}}-\Omega^{\bpsi}_{\m\n}\var{\Act}{\Omega^{\psi}_{\a\b}}-\Omega^{\bpsi}_{\a\b}\var{\Act}{\Omega^{\psi}_{\m\n}}\right)=0.
\end{align}
Detailed investigation of the above relations, and their application to renormalization are in progress and will constitute a major part of a forthcoming paper.
%
\subsection{Propagators:}
%
From the action \eqref{complete-action1} and \eqref{eq:act_complete} with $J/\bar J$ and $Q/\bar Q$ set to their physical values given by \eqref{JQ-phys} one can easily derive the propagators
{\allowdisplaybreaks
\begin{subequations}
\begin{align}
G^{\bc c}(k)&=-\inv{k^2}\,,\label{eq:prop_cc}\\
G^{\bpsi\psi}_{\m\n,\r\s}(k)&=-\frac{\left(\d_{\m\r}\d_{\n\s}-\d_{\m\s}\d_{\n\r}\right)}{2\mu^2k^2\k^2}\,,\label{eq:prop_psipsi}\\
G^{AA}_{\m\n}(k)&=\inv{\nok}\left(\d_{\m\n}-\frac{k_\m k_\n}{k^2}\right)\,,\label{eq:prop_aa}\\
G^{AB}_{\m,\r\s}(k)&=\frac{\ri a'}{2\mu}\frac{\left(k_\r\d_{\m\s}-k_\s\d_{\m\r}\right)}{k^2\k^2\nok}=G^{A\bB}_{\m,\r\s}(k)=-G^{\bB A}_{\r\s,\m}(k)\,,\label{eq:prop_ab}\\
G^{\bB B}_{\m\n,\r\s}(k)&=\frac{-1}{2\mu^2k^2\k^2}\left[\d_{\m\r}\d_{\n\s}-\d_{\m\s}\d_{\n\r}-a'^2\frac{k_\m k_\r\d_{\n\s}+k_\n k_\s\d_{\m\r}-k_\m k_\s\d_{\n\r}-k_\n k_\r\d_{\m\s}}{2k^2\k^2\nok}\right],\label{eq:prop_bbarb}\\
G^{BB}_{\m\n,\r\s}(k)&=\frac{a'^2}{4k^2\k^2}\left[\frac{k_\m k_\r\d_{\n\s}+k_\n k_\s\d_{\m\r}-k_\m k_\s\d_{\n\r}-k_\n k_\r\d_{\m\s}}{\mu^2k^2\k^2\nok}\right]=G^{\bB\bB}_{\m\n,\r\s}(k)\,,\label{eq:prop_bb}
\end{align}
\end{subequations}
where the abbreviation $a'\equiv\l/\mu$ was used.
}
From the form of these propagators we notice that both $G^{B\bB}$ and $G^{\bpsi\psi}$ scale with $1/k^2\k^2$ in the ultraviolet. Furthermore, all vertices with one $B$, one $\bB$ and an arbitrary number of $A$ legs have exactly the same form as the ones with one $\psi$, one $\bpsi$ and an arbitrary number of $A$ legs. Therefore, considering our previous results of~\cite{Blaschke:2009a}, we expect all divergent contributions to the vacuum polarization coming from the $\psi$ sector to exactly cancel those coming from the $B$ sector. Of course this conjecture has to be proven by explicit calculations which are in progress.

Note that the propagators obey the following symmetries and relations:
{\allowdisplaybreaks
\begin{subequations}
\begin{align}
 G^{AB}_{\m,\r\s}(k)&=G^{A\bB}_{\m,\r\s}(k)=-G^{BA}_{\r\s,\m}(k)=-G^{\bB A}_{\r\s,\m}(k),\\
 G^{\phi}_{\m\n,\r\s}(k)&=-G^{\phi}_{\n\m,\r\s}=-G^{\phi}_{\m\n,\s\r}(k)=G^{\phi}_{\n\m,\s\r}(k),\\
 \text{for }\phi&\mathrel{\in}\{\bpsi\psi,\bB B, BB, \bB \bB\},\nonumber\\
 2k^2\k^2 G^{AB}_{\r,\m\n}(k)&=\ri\frac{a'}{\m}\left(k_\m G^{AA}_{\r\n}(k)-k_\n G^{AA}_{\r\m}(k)\right),\label{prop-rel_d}\\
  \inv{\m^2}\left(\d_{\mu\rho}\d_{\nu\s}-\d_{\mu\s}\d_{\nu\rho}\right)&=\ri\frac{a'}{\m}\left(k_\m G^{BA}_{\r\s,\n}(k)-k_\n G^{BA}_{\r\s,\m}(k)\right)-2k^2\k^2 G^{B\bB}_{\mu\nu,\rho\s}(k),\\
  0&=\ri\frac{a'}{\m}\left(k_\m G^{BA}_{\r\s,\n}(k)-k_\n G^{BA}_{\r\s,\m}(k)\right)-2k^2\k^2G^{BB}_{\mu\nu,\rho\s}(k),\label{prop-rel_f}\\
  G^{B\bB}_{\m\n,\r\s}(k)&=G^{\bpsi\psi}_{\m\n,\r\s}(k)+G^{B\bB}_{\m\n,\r\s}(k).
\end{align}
\end{subequations}
In fact, relations \eqref{prop-rel_d}-\eqref{prop-rel_f} follow directly from the equations of motion for $B_{\m\n}$ and $\bB_{\m\n}$.}
%
\subsection{Power counting}
%
\label{sec:power_counting}
The superficial degree of UV divergence is determined by the number of external legs of the various fields denoted by $E$. Its explicit form is given by:
\begin{subequations}
\begin{align}
d_{\g}&=4-E_A-E_{c/\bc}-2E_B-2E_{\bB}-2E_{\psi\bpsi}-2E_{\th}\,,\\
d_{\g}&=4-E_A-E_{c/\bc}-2E_{a'}\,,
\end{align}
\end{subequations}
where
\begin{align}
E_{a'}&=E_B+E_{\bB}+E_{\psi/\bpsi}+E_{\th}\,,
\end{align}
and $E_\th$ counts negative powers of $\th$. It is possible that $E_{\th}$ becomes negative. Hence, the first version (counting $E_{a'}$, i.e. the overall powers of $a'$ in a graph) is probably more useful, as $E_{a'}\geq0$.

%
\section{Conclusion and Outlook}
%
\label{sec:conclusion}
Motivated by the recent work of Vilar \etal~\cite{Vilar:2009} we have proposed a modified version of our translation invariant {\nc} $\inv{p^2}$ U(1) gauge model~\cite{Blaschke:2009a} being localized by introducing auxiliary fields together with associated ghosts and antighosts. This procedure avoids the occurrence of additional degrees of freedom which has been subject of criticism~\cite{Vilar:2009}. The $\inv{p^2}$ model features an IR damping mechanism which suppresses the infamous UV/IR mixing problem of {\nc} QFT (c.f. references \cite{Gurau:2009,Blaschke:2008b,Blaschke:2009a}). Careful construction of BRST doublet structures paves the path for the application of the well known algebraic renormalization procedure~\cite{Piguet:1995}. The unavoidable soft breaking which spoils BRST invariance can be cured in the UV by introducing a doublet of sources to restore the symmetry. Demanding these sources to take appropriate physical values, i.e. to reduce to some unity-valued tensor structure or to vanish (cf. \eqnref{JQ-phys}), one is able to renormalize the high energy behaviour in a standard way while maintaining the desirable IR damping. We have shown that the resulting model obeys the Slavnov-Taylor identity, gauge fixing condition, ghost equation, and the symmetries $U^{(0)}$, $\tilde{U}^{(0)}$, $U^{(1)}$, $\tilde{U}^{(1)}$, and $U^{(2)}$ introduced in Ref.~\cite{Vilar:2009}.\\
It remains to derive the most general counterterm and conduct the loop calculations necessary for the determination of coefficients which is postponed to future publications.

\section*{Acknowledgments}
The work of D.~N.~Blaschke, A.~Rofner and R.~I.~P.~Sedmik was supported by the ``Fonds zur F\"orderung der Wissenschaftlichen Forschung'' (FWF) under contracts P20507-N1 and P19513-N16.
\\
The authors are indebted to F.~Gieres for helpful comments.



\begin{thebibliography}{10}
\expandafter\ifx\csname url\endcsname\relax
  \def\url#1{{\tt #1}}\fi
\expandafter\ifx\csname urlprefix\endcsname\relax\def\urlprefix{URL }\fi
\providecommand{\eprint}[2][]{\url{#2}}

\bibitem{Vilar:2009}
L.~C.~Q. Vilar, O.~S. Ventura, D.~G. Tedesco and V.~E.~R. Lemes, {\it
  Renormalizable {N}oncommutative {U(1)} {G}auge {T}heory {W}ithout {IR/UV}
  {M}ixing\/}, \href{http://www.arxiv.org/abs/0902.2956}{{\tt
  [arXiv:0902.2956]}}.

\bibitem{Blaschke:2009a}
D.~N. Blaschke, A.~Rofner, M.~Schweda and R.~I.~P. Sedmik, {\it {One-Loop
  Calculations for a Translation Invariant Non-Commutative Gauge Model}\/},
  \href{http://www.arxiv.org/abs/0901.1681}{{\tt [arXiv:0901.1681]}}.

\bibitem{Tanasa:2008d}
A.~Tanasa, {\it {Scalar and gauge translation-invariant noncommutative
  models}\/}, \href{http://www.arxiv.org/abs/0808.3703}{{\tt
  [arXiv: 0808.3703]}}.

\bibitem{Rivasseau:2007a}
V.~Rivasseau, {\it Non-commutative renormalization,\/} {in {\it Quantum Spaces
  --- {Poincar\'{e}} Seminar 2007}, B. {Duplantier} and {V. Rivasseau} eds.,
  Birkh\"{a}user Verlag}, \href{http://www.arxiv.org/abs/0705.0705}{{\tt
  [arXiv: 0705.0705]}}.

\bibitem{Douglas:2001}
M.~R. Douglas and N.~A. Nekrasov, {\it Noncommutative field theory\/}, {\it
  Rev. Mod. Phys.\/} {\bf 73} (2001) 977--1029,
  \href{http://www.arxiv.org/abs/hep-th/0106048}{{\tt [arXiv:hep-th/0106048]}}.

\bibitem{Grosse:2003}
H.~Grosse and R.~Wulkenhaar, {\it Renormalisation of {phi**4} theory on
  noncommutative {R**2} in the matrix base\/}, {\it JHEP\/} {\bf 12} (2003)
  019, \href{http://www.arxiv.org/abs/hep-th/0307017}{{\tt
  [arXiv:hep-th/0307017]}}.

\bibitem{Grosse:2008a}
H.~Grosse and F.~Vignes-Tourneret, {\it Minimalist translation-invariant
  non-commu\-ta\-tive scalar field theory\/},
  \href{http://www.arxiv.org/abs/0803.1035v1}{{\tt [arXiv:0803.1035v1]}}.

\bibitem{Rivasseau:2005a}
V.~Rivasseau, F.~Vignes-Tourneret and R.~Wulkenhaar, {\it Renormalization of
  noncommutative {phi**4}-theory by multi-scale analysis\/}, {\it Commun. Math.
  Phys.\/} {\bf 262} (2006) 565--594,
  \href{http://www.arxiv.org/abs/hep-th/0501036}{{\tt [arXiv:hep-th/0501036]}}.

\bibitem{Rivasseau:2005b}
R.~Gurau, J.~Magnen, V.~Rivasseau and F.~Vignes-Tourneret, {\it Renormalization
  of non-commutative {phi**4(4)} field theory in x space\/}, {\it Commun. Math.
  Phys.\/} {\bf 267} (2006) 515--542,
  \href{http://www.arxiv.org/abs/hep-th/0512271}{{\tt [arXiv:hep-th/0512271]}}.

\bibitem{Grosse:2004b}
H.~Grosse and R.~Wulkenhaar, {\it Renormalisation of {phi**4} theory on
  noncommutative {R**4} in the matrix base\/}, {\it Commun. Math. Phys.\/} {\bf
  256} (2005) 305--374, \href{http://www.arxiv.org/abs/hep-th/0401128}{{\tt
  [arXiv:hep-th/0401128]}}.

\bibitem{Gurau:2009}
R.~Gurau, J.~Magnen, V.~Rivasseau and A.~Tanasa, {\it A translation-invariant
  renormalizable non-commutative scalar model\/}, {\it Commun. Math. Phys.\/}
  {\bf 287} (2009) 275--290, \href{http://www.arxiv.org/abs/0802.0791}{{\tt
  [arXiv:0802.0791]}}.

\bibitem{Blaschke:2008b}
D.~N. Blaschke, F.~Gieres, E.~Kronberger, T.~Reis, M.~Schweda and R.~I.~P.
  Sedmik, {\it Quantum Corrections for Translation-Invariant Renormalizable
  Non-Commutative $\Phi^4$ Theory\/}, {\it JHEP\/} {\bf 11} (2008) 074,
  \href{http://www.arxiv.org/abs/0807.3270}{{\tt [arXiv:0807.3270]}}.

\bibitem{Tanasa:2008a}
J.~B. Geloun and A.~Tanasa, {\it {One-loop $\beta$ functions of a
  translation-invariant renormalizable noncommutative scalar model}\/}, {\it
  Lett. Math. Phys.\/} {\bf 86} (2008) 19--32,
  \href{http://www.arxiv.org/abs/0806.3886}{{\tt [arXiv:0806.3886]}}.

\bibitem{Blaschke:2008a}
D.~N. Blaschke, F.~Gieres, E.~Kronberger, M.~Schweda and M.~Wohlgenannt, {\it
  {Translation-invariant models for non-commutative gauge fields}\/}, {\it J.
  Phys.\/} {\bf A41} (2008) 252002,
  \href{http://www.arxiv.org/abs/0804.1914}{{\tt [arXiv:0804.1914]}}.

\bibitem{Zwanziger:1989}
D.~Zwanziger, {\it {Local and Renormalizable Action from the {Gribov}
  Horizon}\/}, {\it Nucl. Phys.\/} {\bf B323} (1989) 513--544.

\bibitem{Dudal:2008}
D.~Dudal, J.~Gracey, S.~P. Sorella, N.~Vandersickel and H.~Verschelde, {\it A
  refinement of the Gribov-Zwanziger approach in the Landau gauge: infrared
  propagators in harmony with the lattice results\/}, {\it Physical Review D\/}
  {\bf 78} (2008) 065047, \href{http://www.arxiv.org/abs/0806.4348}{{\tt
  [arXiv:0806.4348]}}.

\bibitem{Baulieu:2009}
L.~Baulieu and S.~P. Sorella, {\it Soft breaking of {BRST} invariance for
  introducing non-perturbative infrared effects in a local and renormalizable
  way\/}, {\it Physics Letters B\/} {\bf 671} (2009) 481,
  \href{http://www.arxiv.org/abs/0808.1356}{{\tt [arXiv:0808.1356]}}.

\bibitem{Zwanziger:1993}
D.~Zwanziger, {\it {Renormalizability of the critical limit of lattice gauge
  theory by {BRS} invariance}\/}, {\it Nucl. Phys.\/} {\bf B399} (1993)
  477--513.

\bibitem{Piguet:1995}
O.~Piguet and S.~P. Sorella, {\it Algebraic renormalization: Perturbative
  renormalization, symmetries and anomalies\/}, {\it Lect. Notes Phys.\/} {\bf
  M28} (1995) 1--134.

\end{thebibliography}

\end{document}